# SYNCHROTRON RADIATION DAMPING, INTRABEAM SCATTERING AND BEAM-BEAM SIMULATIONS FOR HE-LHC

A. Valishev, Fermilab, Batavia, IL 60510, U.S.A.

*Abstract*

The proposed High-Energy LHC project presents an unusual combination of strong synchrotron radiation (SR) damping and intrabeam scattering (IBS), which is not seen in present-day hadron colliders. The subject of investigation reported in this paper was the simulation of beam-beam effect for the HE-LHC parameters. Parameters of SR and IBS are calculated, and the luminosity evolution is simulated in the absence of beam-beam interaction. Then, a weak-strong numerical simulation is used to predict the effect of beam-beam interaction on particle losses and emittance evolution.

## MACHINE AND BEAM PARAMETERS

Main parameters of HE-LHC relevant for our calculations are presented in Table 1.

Table 1: Machine and beam parameters

| Parameter | Value |
|---|---|
| Beam energy | 16.5 TeV |
| Number of bunches | 1404 |
| Number of interaction points | 2 |
| Bunch population | $1.3 \times 10^{11}$ |
| Initial normalized transverse emittance | 3.75, 1.84 (x,y) $\mu$m |
| Initial momentum spread | $0.9 \times 10^{-4}$ |
| RF voltage | 32 MV |
| Beta-function at IP | 1.0, 0.43 (x,y) m |
| Full crossing angle | 175 $\mu$rad |

## PARAMETERS OF SYNCHROTRON RADIATION

Calculation of synchrotron radiation integrals was based on current LHC optics V6.5. Main parameters of SR and equilibrium emittance were derived using the conventional formulae [1] (see Table 2):

$$U_0 = \frac{C_\gamma}{2\pi} E^4 I_2$$

$$\tau_{x,y} = \frac{ET_0}{U_0}, \tau_E = \frac{ET_0}{2 \cdot U_0}$$

$$\frac{d\varepsilon_x}{dt} = -\frac{\varepsilon_x}{\tau_x} + \frac{55}{48\sqrt{3}} \frac{\hbar c}{T_0} \frac{r_0}{mc^2} \gamma^5 I_5$$

---

Table 2: Synchrotron radiation parameters

| Parameter | Value |
|---|---|
| Synchrotron radiation integrals | $I_2$=0.002245 m$^{-1}$ |
|  | $I_3$=7.99$\times 10^{-7}$ m$^{-2}$ |
|  | $I_5$=2.11$\times 10^{-8}$ m$^{-1}$ |
| Energy loss per turn | $U_0$=206.3 keV |
| SR power | 67 kW |
| Emittance damping time | $\tau_x$, $\tau_y$=1.93 h |
|  | $\tau_E$=0.96 h |
| Normalized equilibrium emittance | 0.01 $\mu$m |
| Equilibrium momentum spread | $3.4 \times 10^{-6}$ |

Note that the equilibrium emittance and momentum spread due to synchrotron radiation are much smaller than the initial values, and the damping time is significantly shorter than the expected store duration (10 hours).

## INTRA-BEAM SCATTERING AND LUMINOSITY EVOLUTION

Intrabeam scattering was treated in the smooth optics approximation by V.Lebedev [2] with the main parameters listed in Table 3.

$$\frac{d}{dt}\begin{pmatrix} \varepsilon_x \\ \varepsilon_y \\ \sigma_E^2 \end{pmatrix} = \frac{N r_0^2 c L_c}{4\sqrt{2} \beta^3 \gamma^3 \sigma_x \sigma_y \sigma_z \theta_\perp} \begin{pmatrix} \langle A_x \rangle_s \\ 0 \\ 1 \end{pmatrix}$$

Table 3: Parameters of IBS

| Parameter | Value |
|---|---|
| Lattice parameters (LHC V6.5) | $\langle \beta_x \rangle$=104.8 m |
|  | $\langle \beta_y \rangle$=109.4 m |
|  | $\langle A_x \rangle$=2.29 |
| Horizontal emittance growth time | 82 h |
| Longitudinal emittance growth time | 72 h |

Evolution of the beam parameters and luminosity was then calculated via the numerical solution of the following system of equations:

$$\frac{d\varepsilon_x}{dt} = -\frac{2\varepsilon_x}{\tau_{SRx}} + \frac{d\varepsilon_{xSR}}{dt} + \frac{d\varepsilon_{xIBS}}{dt} + \frac{d\varepsilon_{xBB}}{dt}$$

$$\frac{d\varepsilon_y}{dt} = -\frac{2\varepsilon_y}{\tau_{SRy}} + \frac{d\varepsilon_{ySR}}{dt} + \frac{d\varepsilon_{yIBS}}{dt} + \frac{d\varepsilon_{yBB}}{dt}$$

$$\frac{d\sigma_E^2}{dt} = -\frac{2\sigma_E^2}{\tau_{SRE}} + \frac{d\sigma_{SR}^2}{dt} + \frac{d\sigma_{IBS}^2}{dt} + \frac{d\sigma_{BB}^2}{dt}$$

$$\frac{dN}{dt} = -N_{IP}\frac{L}{N_b}\sigma_{tot}$$

Here $L$ is the luminosity, $N_b$ is the number of bunches, $\sigma_{tot}$ is the p-p interaction cross-section, $N_{IP}$ is the number of IPs, indices SR, IBS and BB label the emittance growth (or damping) rates for synchrotron radiation, intrabeam scattering, and scattering at IPs, respectively. These growth rates are calculated at every step of numerical integration for current beam parameters.

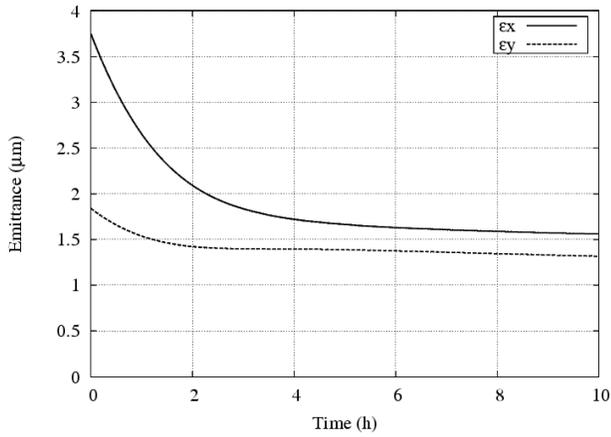

Figure 1: Evolution of transverse emittances.

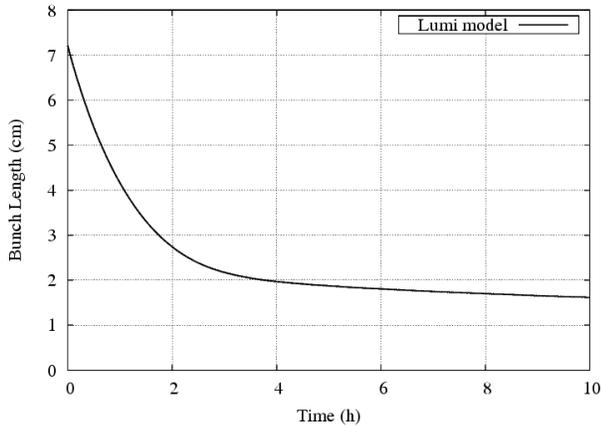

Figure 2: Evolution of bunch length.

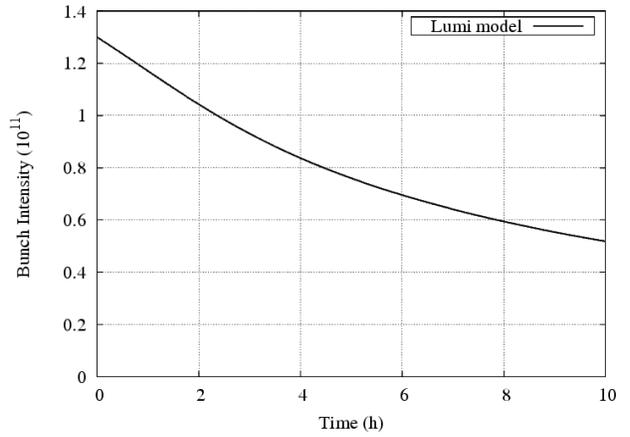

Figure 3: Evolution of bunch intensity.

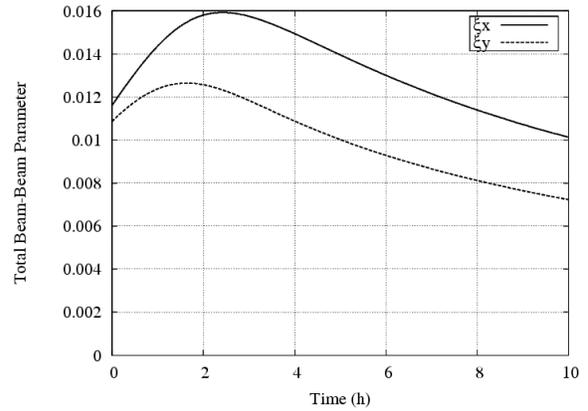

Figure 4: Beam-beam parameters vs. time.

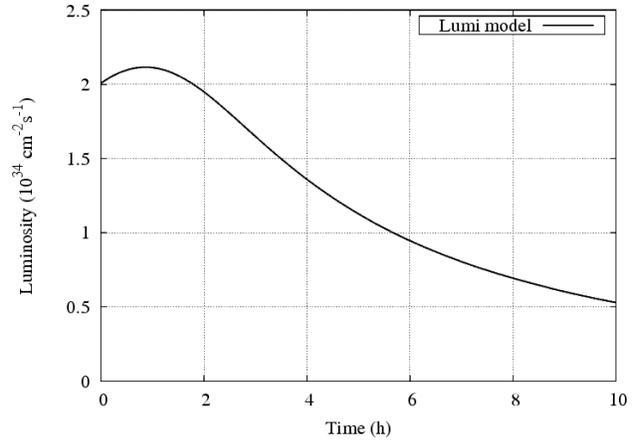

Figure 5: Evolution of luminosity. Luminosity integral over 10 h is 450 pb$^{-1}$.

Figures 1-5 present the calculated evolution of beam parameters over a period of 10 hours. An important feature of the beam dynamics is the reduction of transverse emittances due to synchrotron radiation by approximately a factor of two. Combined with the beam intensity decay caused by luminous particle losses, this results in the increase of the total beam-beam parameter by 50% (from the initial value of 0.01 to 0.016).

## BEAM-BEAM SIMULATION

Numerical simulation was performed with the use of weak-strong tracking code Lifetrac. A bunch of 5,000 macro-particles with weighted Gaussian initial distribution was tracked for $5\times10^7$ turns (which corresponds to 1.2 h of beam time) in order to evaluate the importance of beam-beam effects. The machine optics was represented by linear 6D maps, no long-range collisions were considered. The synchrotron radiation damping, quantum excitation and intrabeam scattering were represented by kicks applied once per turn. The code does not include particle losses due to luminosity and diffusion caused by scattering at the IPs.

Figures 6-8 present the results of numerical simulation along with the curves obtained using the luminosity evolution model described in the previous section. Numerical simulation did not produce any particle losses. One can see that beam-beam interaction does not cause additional emittance growth.

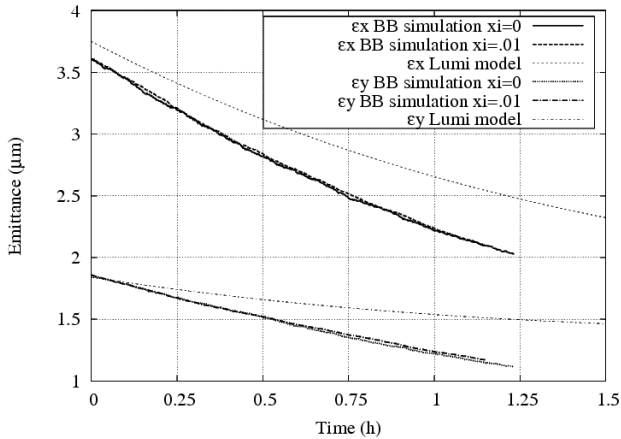

Figure 6: Beam emittances vs. time. Comparison of beam-beam simulation and luminosity model.

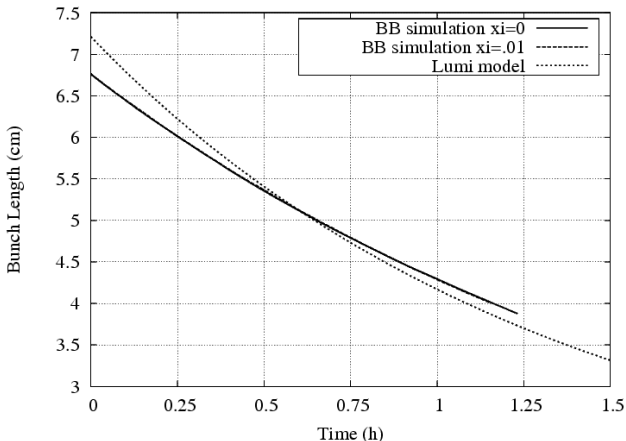

Figure 7: Comparison of bunch length from numerical beam-beam simulation and luminosity model.

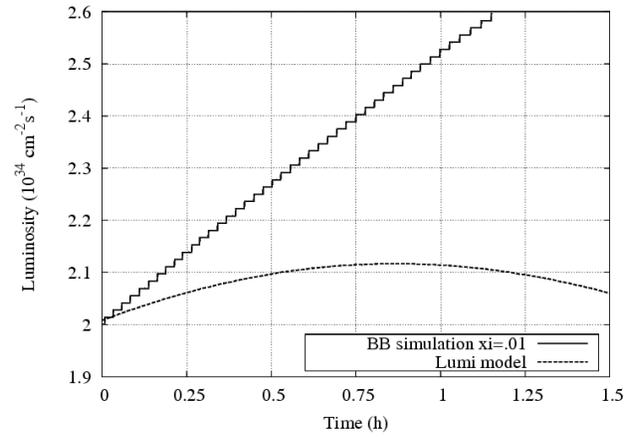

Figure 8: Luminosity vs. time. Beam-beam simulation and luminosity evolution model. Numerical curve is higher owing to the absence of the luminous beam decay in the model.

## SUMMARY

Combination of parameters of the proposed High Energy LHC produces beam dynamics not observed at present hadron colliders. Synchrotron radiation causes emittance damping with the characteristic time of 2 hours. Intrabeam scattering, typically the dominant effect, is relatively weak with the initial emittance growth rate of 70 hours. As the beam emittance decreases, synchrotron radiation and intrabeam scattering come to equilibrium, a situation typical for low emittance electron damping rings. The resulting beam emittance is approximately half of the initial value. As the result, the beam-beam parameter experiences growth over the initial 2 hours of the store.

Numerical simulations with a weak-strong particle tracking code, which included major effects, predict that beam-beam effects would cause no particle losses. The evolution of beam emittance is not modified by beam-beam interaction. Modeling confirms that the luminosity integral of 450 pb$^{-1}$ over 10 hours is achievable. The simulations employed a simplified accelerator model, in which no nonlinearities existed. Hence it is reasonable to expect extra losses due to machine nonlinearities.